\documentstyle[12pt,twoside,fleqn,espcrc1]{article}

\newcommand{\bce}{\begin{center}}
\newcommand{\ece}{\end{center}}
\newcommand{\bea}{\begin{eqnarray}}
\newcommand{\eea}{\end{eqnarray}}
\newcommand{\be}{\begin{equation}}
\newcommand{\ee}{\end{equation}}
\newcommand{\bd}{\begin{displaymath}}
\newcommand{\ed}{\end{displaymath}}
\newcommand{\bit}{\begin{itemize}}
\newcommand{\eit}{\end{itemize}}
\newcommand{\ben}{\begin{enumerate}}
\newcommand{\een}{\end{enumerate}}
\newcommand{\bdes}{\begin{description}}
\newcommand{\edes}{\end{description}}
\newcommand{\dslash}{\partial \hspace{-6pt}/}
\newcommand{\Aslash}{A \hspace{-6pt}/}

\newcommand{\AmS}{{\protect\the\textfont2
  A\kern-.1667em\lower.5ex\hbox{M}\kern-.125emS}}

\title{Variational treatment of quenched QED using the worldline
technique }

\author{A.~W.~Schreiber\address{Department of Physics and Mathematical
Physics and 
Research Centre for the Subatomic Structure of Matter, University of
Adelaide, 
Adelaide, S. A. 5005, Australia }, 
C.~Alexandrou\address{Department of Natural Sciences, University of
Cyprus, 
CY-1678 Nicosia, Cyprus} and 
R.~Rosenfelder\address{Paul Scherrer Institute, CH-5232 Villigen PSI, 
Switzerland}}

\begin{document}
\maketitle
\vspace{-8cm}
\begin{flushright}
{\footnotesize 
Invited talk at the 15th Int. Conf. on Few-Body Problems in Physics \\
Groningen, The Netherlands, 22-26 July, 1997 \\
ADP-97-35-T267, UCY-PHY-97/03}
\end{flushright}
\vspace{6cm}



\section{INTRODUCTION}

Perturbative Quantum Electrodynamics has been successfully tested to 
unparalleled precision for around half a century.  In addition, and
rather more
recently, the theory's behaviour in its strong coupling limit has
attracted
a growing amount of attention.  The reason for this is not only that it
serves as a simple prototype gauge theory for testing nonperturbative
calculational techniques
which one then goes on to apply to QCD, but also because it 
exhibits interesting behaviour in its own right.  In particular, it now
appears
to be quite well established that QED breaks chiral symmetry dynamically 
as long as the coupling is sufficiently large.

Traditionally,  strongly coupled QED has been studied using
Schwinger-Dyson,
Lattice and renormalization group approaches.  In this contribution we 
describe a rather different method, namely a variational approach 
 using the worldline formulation of field theory. We have
developed this technique in the context of a scalar field
theory~\cite{scalar}
and we applied it here, for the first time, to a gauge theory with 
spin-${1 \over 2}$ particles.
 Although it is
too early for quantitative results, qualitatively the approach seems
promising: The method is formulated covariantly in Minkowski
space-time, the Ward-Takahashi
identities are satisfied, the connection between the bare- and
pole-masses of the electron is gauge independent, the known exact
result for the infrared dependence of the on-shell renormalized
propagator is reproduced and corrections to the variational
approximation can
be calculated systematically.

\section{THE WORLDLINE TECHNIQUE}
In recent years a rather large literature on the worldline technique has 
developed~\cite{Schub}, so it suffices here to just provide a general
outline.
  We shall concentrate on Green functions within quenched QED involving
only one fermion line -- i.e. the propagator, vertex function, Compton
scattering etc.  The theory is quenched by integrating out the
fermions and setting the resulting determinant equal to unity.  After
differentiating the appropriate generating functional $Z$ with respect
to 
the external fermion's
source one obtains the generating functional for the general $(2+n)$-point
function
\be
Z'\> [j,y] 
 =  \int {\cal D}A \>  <y \> |\> \frac{1}{i\dslash-g\Aslash-M_0}\> |\>0> 
\> \exp \Bigl \{ i S_0 [ A ] + (j,A) \> \Bigr \} \> .
\label{gen' functional}
\ee
Here $j$ is the source for the the photon field $A$, $M_0$ is the bare
mass
of the electron, $g$ is the coupling constant and $S_0[ A ]$ is the
gauge part of the standard QED action with Lagrangian
\be
{\cal L}_0 (A)  = -\frac{1}{4} F_{\mu\nu} F^{\mu\nu} + \frac{1}{2} 
m_{\gamma}^2 A^2 
-\frac{1}{2 \xi} (\partial \cdot A)^2 \> .
\ee
In order to regulate infrared divergences 
 a photon mass $m_{\gamma}$ has been introduced 
and the gauge has been fixed covariantly by employing the
gauge fixing parameter $\xi$. 

We would like to integrate out the photon field. It is not possible to 
do this at this stage because it appears in the electron 
propagator in Eq.~(\ref{gen' functional}) in a non-gaussian way.
In order to proceed, the fermion is re-introduced in terms of a path
integral 
over its
worldline $x(t)$ as well as a corresponding path integral over a
Grassmann
valued function $\zeta(t)$ which contains the information about the
electron's
spin degree of freedom~\cite{FrGi}.  One obtains
\be
<y\> |\> \frac{1}{i\dslash-g\Aslash-M_0}\> |\> 0> \> = \>
\int_0^{\infty} dT \> N(T) \>  \int d\chi \int{\cal D}x {\cal D}\zeta
 \> e^{  \> i S [x,\zeta] }
\;\;\;,
\label{G hat propagator}
\ee
where $N(T)$
provides the proper normalization, the integration
parameter $\chi$ is a Grassmannian partner
to the proper time $T$, the four-dimensional bosonic path integral has
the 
boundary conditions $x(0) = 0$ and $x(T) = y$ while the five dimensional
fermionic path integral has the single boundary condition
$\zeta^\mu(0) + \zeta^\mu(T) = i \gamma^5 \gamma^\mu$, 
$\zeta^5(0) + \zeta^5(T) = i \gamma^5 $.  
The worldline action $S [x,\zeta]$ is given by
\be
S[x,\zeta] \> = \> S_0[x,\zeta] \> + \> g \int_0^T dt \> \Bigl \{ \> 
 -  \dot{x}_{\mu}(t)\> A^{\mu}[x(t)]
\>+\> i\>  F_{\mu\nu}[x(t)]\>\zeta^{\mu}(t)\>\zeta^{\nu}(t) \> \Bigr \}
\;\;\;.
\label{eq: worldline action}
\ee
  As expected for a relativistic 
particle the free part of the action $S_0[x,\zeta]$ includes, apart
from a kinetic term, further contributions
which provide coupling between orbital and 
spin motion as well as a kinetic term for the spin
\be
S_0[x,\zeta] \! = \! \! \int_0^T \! \! \! dt  \left \{-{ \dot{x}^2(t) \over 2}
 - {i \over T} \left[ \dot{x}^{\mu}(t)\zeta_{\mu}(t)
- M_0 \zeta^5(t)\right]\chi 
 - i \zeta_m(t) \dot \zeta^m(t) - {i \over T}\zeta_m(T) \zeta^m(0) 
 \right \}.
\label{eq: s0}
\ee
Here $ m = 0,1,2,3,5 $ and the metric $ \> g^{m n} = {\rm diag}(1,-1,-1,-1,-1) 
\> $ is used for short-hand notation.
Note that the gauge field $A$ appears linearly in 
Eq.~(\ref{eq: worldline action}) so that we may now
integrate out the photons in the generating functional 
~(\ref{gen' functional}).  Differentiating $n$ times with
respect to the photon source $j$ (and then setting $j=0$)
as well as taking the Fourier transform yields the
general Green function of an electron interacting with $n$
external photons.
With $p$ defined as the ingoing electron momentum and $p'$ ($\{q_i\}$)
the
outgoing electron (photon) momenta, and the external photon legs as well
as the momentum conserving delta function removed, one obtains
\bea
G_{2,n}^{\{\mu_i\}} \left (p,p';\{q_i\}\right ) &=& {\rm const. } \> 
\int_0^{\infty} \! dT \int d\chi  \> 
\exp \left[  \> - {i \over 2} M_0^2 T \> + \> {i \over 2} M_0 \gamma_5 \chi
\> \right ]  \>  
\int {\cal D} \tilde x {\cal D} \zeta \>  e^{i [ S_{\rm eff} + p' \cdot
y]}\nonumber \\
&& \hspace{2cm} \cdot \prod_{i=1}^{n} \left [g
\int_0^T \> d \tau_i \>e^{i q_i \cdot x(\tau_i)} \>
J^{\mu_i} [ -q_i, x(\tau_i),\zeta(\tau_i)]\right ]\> \gamma_5\;\;\;,
\label{eq: green function}
\eea
where the ${\cal D} \tilde x$ implies an integration over the endpoint
$y$ 
in addition to the path integral ${\cal D} x$
(similarly, we will henceforth define $\tilde S_{\rm eff}$ to be 
$S_{\rm eff} + p' \cdot y$).
Because the 5$^{\rm th}$ component of the Grassmann function $\zeta$ is
non-interacting, this integration has been carried out in this equation. 
The effective action  contains the free term in Eq.~(\ref{eq: s0}) as well
as an 
interaction dependent piece
\bea
S_I[x,\zeta] &=& - \frac{g^2}{2} \int_0^T dt_1 dt_2 \int \frac{d^4 q}{(2
\pi)^4}
 \> G^{\mu\nu}(q)  \> J_{\mu} \left [ q,x(t_1),\zeta(t_1) \right ] \> 
J_{\nu} \left [ - q,x(t_2),\zeta(t_2) \right ] \nonumber \\
&& \hspace{7cm} \cdot \> \exp \left [ \> - i q \cdot ( x(t_1) - x(t_2) ) \> 
\right ] \> .
\label{S1} 
\eea
Here $G^{\mu\nu}$ is the photon propagator and the fermion current is 
\be
J_{\mu} \left [ q,x(t),\zeta(t) \right ] \> = \> \dot x_{\mu}(t) \> + \> 
2  \, \zeta_{\mu}(t) \> q \cdot \zeta(t)\;\;\;,
\label{current}
\ee
which corresponds to the usual Gordon decomposition
for the Dirac current of a particle of unit mass 
into a convection and a spin current.

It is a nice feature of the worldline representation that it is trivial
to verify the
Ward-Takahashi identities for an arbitrary Green function.
Because of the Grassmannian nature of $\zeta(t)$ a contraction of
an external momentum $q^\mu$ into the appropriate current
$J_{\mu} $ just yields a total
derivative, i.e. 
$q_i \cdot  J_i  \> = \> q_i \cdot \dot x(\tau_i)$,
which allows one to do the $\tau_i$ integration associated with  $q_i$,
 i.e.
\be
\int_0^T \> d \tau_i \>e^{i q_i \cdot x(\tau_i)} \>
q_i \cdot J [- q_i, x(\tau_i),\zeta(\tau_i)] \> = \>
{1 \over i}\left ( e^{i q_i \cdot y} \> - \> 1 \right )\;\;\;.
\label{eq: tau int}
\ee
The remaining $q_i$ dependence here may be absorbed in a shifted
momentum $p'$ in the momentum space action $S_{\rm eff}  + p' \cdot y$ and
so $q_{\mu_j} \cdot G_{2,n}^{\{\mu_i\}}$ is related to the difference
of two (2+$(n-1))$-point functions:
\be
{i \over g}\> q_{\mu_j} \cdot G_{2,n}^{\{\mu_i\}}(p,p';\{q_i\}) \> = \>
G_{2,n-1}^{\{\mu_i\}}(p+q_i,p'+q_i;\{q_i\})\> - \>
G_{2,n-1}^{\{\mu_i\}}(p,p';\{q_i\})\;\;\;,
\ee
where the notation should be self explanatory.  This is the correct
result for the Ward-Takahashi identity for the  $(2+n)$-point
function untruncated in the external fermion legs.
Because of the simplicity of the above argument,
based only on the structure of the fermionic current,
it is quite
easy to ensure that the variational approximation doesn't violate
it.

\section{THE VARIATIONAL APPROXIMATION}

\vspace*{-0.2cm}
Eq.~(\ref{eq: green function}) provides an exact representation
 of all Green functions occurring in quenched QED involving only one 
fermion line.  It is not possible to carry out the remaining 
path integrals
over $x(t)$ and $\zeta(t)$ because they do not occur quadratically in
the action 
(see Eq.~(\ref{S1})).  We shall calculate the Green functions of the
theory 
in an approximate manner through the use of the Feynman-Jensen 
variational principle.  
This technique has a rather long history, 
having its
origin in the polaron problem~\cite{Feyn} and it was applied, more
recently, to a scalar 
relativistic 
field theory by us~\cite{scalar}.  In order to illustrate the idea, let
us consider the
electron propagator (i. e. $n=0$ in Eq.~(\ref{eq: green function})).
We rewrite the path integral by adding and subtracting to $\tilde S_{\rm
eff}$ a trial action
$\tilde S_t$:
\be
\int {\cal D} \tilde x {\cal D} \zeta \> e^{i  \tilde S_{\rm eff}}
\>=\> \left ( \int {\cal D} \tilde x {\cal D} \zeta \> e^{i  \tilde S_t}
\right )
{\int {\cal D} \tilde x {\cal D} \zeta  \> e^{i  (\tilde S_{\rm eff}
- \tilde S_t)} \> e^{i  \tilde S_t} \over
\int {\cal D} \tilde x {\cal D} \zeta   \> e^{i  \tilde S_t} } 
\>\equiv\> \left ( \int {\cal D} \tilde x {\cal D} \zeta \> e^{i  \tilde
S_t} \right ) \> \langle  e^{i  \Delta \tilde S}\rangle_{\tilde
S_t}\;\;\;,
\label{eq: rewrite}
\ee
i.e. we need to evaluate the average of the exponential 
of $i  \Delta \tilde S$, weighted with the function 
$e^{i  \tilde S_t}$.
The approximation consists of replacing the average of the exponential
by the
exponential of the average, i.e.
\be
\langle  e^{i  \Delta \tilde S}\rangle_{\tilde S_t} \approx
 e^{i \langle \Delta \tilde S\rangle_{\tilde S_t}}\;\;\;,
\label{eq: approx}
\ee
where $ \> \approx \> $ indicates equality at the stationary point of the right 
hand side 
under unrestricted variations of the functional $\tilde S_t$. 
If one restricts $\tilde S_t$ to depend only on (non-local) quadratic 
functions of $x(t)$ and $\zeta(t)$ then the path integrals may be
performed analytically. However, only
the parameters in the trial action are then left for variation
 and so it is important to note that one can 
systematically calculate the corrections to the variational approximation 
by including higher terms in the 
cumulant expansion of which Eq.~(\ref{eq: approx}) is the first term.
Of course, in order for the approximation to be a useful one, the
essential
physics should already be contained in the leading term.

\section{THE VARIATIONAL PROPAGATOR}
 
In order to assess the feasibility of the program outlined above we have
calculated the electron propagator, using a very simple 
trial action
\be
\tilde S_t = \lambda' p \cdot y \> + \> S_0[x,\zeta] \> + \>
\int_0^T dt_1 dt_2 \> f(t_1-t_2) \> \Bigl [ \> x(t_1) - x(t_2) \> 
\Bigr ] ^2\;\;\;,
\label{trial action}
\ee
i.e. one which has the same dependence on the fermionic degrees of
freedom $\zeta(t)$ as the free action. Only the dependence on the bosonic
degrees
of freedom $x(t)$ has been changed from the free action through the use
of 
the variational parameter $\lambda'$ and the variational function $f$
which describes the retardation effects on the electron due to multiple 
photon emission and re-absorption.
This is certainly not an optimal trial action and 
it only serves to illustrate some of the qualitative features
of what can be expected.  In particular, the trial action (\ref{trial action})
does not respect the supersymmetry between orbital and spin degrees of 
freedom which is present in the exact wordline action \cite{Susy}.
Work with a more general trial action is underway.
  Also, we shall only calculate the
propagator near its mass-shell:  This has the great advantage that in
Fourier space the trial action effectively diagonalizes, i.e.
\be
\int_0^T dt_1 dt_2 f(t_1-t_2) \> \Bigl [ \> x(t_1) - x(t_2) \> \Bigr ]^2 
\>=\> \sum_{k,k'=0}^{\infty} A_{kk'}\> b_k \cdot b_k' 
\hspace{4mm}{\stackrel{\scriptstyle{\rm onshell}}{\Longrightarrow}}
\hspace{4mm}\sum_{k=0}^{\infty} A_{kk}\> b^2_k \;\;\;,
\ee
where $b_k$ are the Fourier  components of the path $x(t)$ and
$A_{kk}$ become the variational parameters taking the place of the
function $ \> f(t_1-t_2) \> $ (not to be confused with the photon field
which has been integrated out).

With this trial action, and employing Pauli-Villars regularization as well
as on-shell renormalization, the connection between the physical mass
$M_{\rm phys}$ and the bare mass $M_0$ becomes 
\be
M_{\rm phys}^2 \left \{ 1 - (1-\lambda)^2 
 -   2 \>{ \Omega[A,\lambda] \> + \> V[A,\lambda] \over M_{\rm
phys}^2}  
\right \}  - M_0^2 = 0
\label{eq: pole position}
\ee
where $ \> \lambda = \lambda' / A_{00} \> $ ( first discussed
by  Mano  \cite{Mano} ).
The variational
parameters should be fixed 
through the variational equation
\be
\delta\left \{ \> M_{\rm phys}^2  \left[ 1 - (1-\lambda)^2 \right ]
 -   2 \>  \Omega[A,\lambda] + \>  V[A,\lambda] \> 
 \right \} = 0\;\;\;.
\label{eq: variational equation}
\ee
Here $V[A,\lambda]$ is a functional which essentially arises from the
required averaging
of $S_I$ (see Eq.~(\ref{eq: rewrite})) while  
$\Omega[A,\lambda]$ gets contributions from the averaging of 
$\tilde S_0 - \tilde S_t$ and the quadratic fluctuations of the trial 
action (apart from the Minkowski formulation it is identical with
the expression given in Ref.~\cite{scalar}).
Both $V$ and $\Omega$ are gauge independent, as they should be, and
hence
the variational parameters are also gauge independent.
If, for $ \> M_0 = 0 \> $, Eqs.~(\ref{eq: pole position}) 
and~(\ref{eq: variational equation}) have a common solution  with
$M_{\rm phys}\neq 0$ then the theory exhibits dynamical mass
generation.
At present it is not known whether the above trial action admits a
solution such as this.  

The contributions to the residue, on the other hand, may conveniently
be broken up into a term $F_g^{\chi}$ which is proportional to $\chi$
and two terms, $F_g^I$  and $F_{qq}^I$, which are not. 
(The subscripts
$g$  and $qq$ on these functions indicate whether they arise from the
$g_{\mu \nu}$ or $q_{\mu} q_{\nu}$ terms in the photon propagator.)
We obtain
\be
Z_2 \> = \> {N_0(A) \over 2}
\left [ \> 1 + { M_0 \over \lambda\> M_{\rm phys}} +  
\frac{1 - \lambda}{\lambda A_{0,0}}
 +  {2  \> i\over \lambda} \> F_g^{\chi}(A,\lambda) \> \right ]
\> e^{ i \>  \left [ \> F_g^I(A,\lambda) +  F_{qq}^I \> \right ]}\;\;\;,
\ee
where $N_0(A)$ is an overall factor of no interest here.  The important
thing to note  is that this expression  only
contains infrared divergences in the functions occurring in the
exponential
factor, and these may in fact be extracted {\it analytically}.  
Indeed, the only gauge dependent function $F_{qq}^I$ has the closed form
\be
F_{qq}^{I} \> = \>- i {\alpha \over 4 \pi} \left ( 1 - \xi \right )
\log {\Lambda^2 \over m_{\gamma}^2} \;\;\;,
\label{eq: fqqI}
\ee
resulting (together with the additional infrared divergence contained in
$F_{g}^I$) in a residue which contains all its infrared and gauge
dependent
behaviour in the overall prefactor 
\be
Z_2\> \propto \> \left ({\Lambda \over m_{\gamma}}\right )^{ {\alpha 
\over 2 \pi}
\left (3 - \xi\right )}\;\;\;.
\ee
This is the well-known Bloch-Nordsieck result \cite{Brown}.

In conclusion, we have summarized here the preliminary results which
have been obtained in the first application of the polaron variational
technique to QED.  We believe that the qualitative success shown so far
warrants a  more quantitative study using a more realistic trial
action.  
Work on this is underway and will be published elsewhere.

\end{document}